\documentclass[aps,twocolumn,superscriptaddress,preprintnumbers,showpacs,article,prb,nobibnotes]{revtex4-2}

\usepackage[latin9]{inputenc}
\usepackage{color}
\usepackage{verbatim}
\usepackage{amsmath}
\usepackage{amssymb}
\usepackage{graphicx}
\usepackage{esint}
\usepackage{esvect}
\usepackage{braket}
\usepackage{amsmath,bm}
\usepackage{graphicx}
\usepackage{subfigure}
\usepackage{url}
\usepackage{lipsum}
\usepackage{lipsum}
\usepackage[dvipsnames]{xcolor}
\usepackage{hyperref}
\usepackage[resetlabels]{multibib}

\newcommand{\bi}{\begin{itemize}}
\newcommand{\ei}{\end{itemize}}
\newcommand{\be}{\begin{enumerate}}
\newcommand{\ee}{\end{enumerate}}

\newenvironment{dfn}{{\vspace*{1ex} \noindent \bf Definition }}{\vspace*{1ex}}

\newcommand{\nn}{\nonumber}  %

	\newcommand{\beq}{\begin{eqnarray}}
	\newcommand{\eeq}{\end{eqnarray}}

\makeatletter
\@ifundefined{textcolor}{}
{%
 \definecolor{BLACK}{gray}{0}
 \definecolor{WHITE}{gray}{1}
 \definecolor{RED}{rgb}{1,0,0}
 \definecolor{GREEN}{rgb}{0,1,0}
 \definecolor{BLUE}{rgb}{0,0,1}
 \definecolor{CYAN}{cmyk}{1,0,0,0}
 \definecolor{MAGENTA}{cmyk}{0,1,0,0}
 \definecolor{YELLOW}{cmyk}{0,0,1,0}
}

\usepackage{mathrsfs}

\draft
\makeatother
\begin{document}

\title{Chiral superconductivity from parent Chern band and its nonabelian generalization}
\author{Yan-Qi Wang}
\thanks{These two authors contributed equally.}
\affiliation{Department of Physics and Joint Quantum Institute, University of Maryland,
College Park, Maryland 20742, USA}
\author{Zhi-Qiang Gao}
\thanks{These two authors contributed equally.}
\affiliation{Department of Physics, University of California,  Berkeley, California 94720, USA}
\affiliation{Materials Sciences Division, Lawrence Berkeley National Laboratory, Berkeley, California 94720, USA}
\author{Hui Yang}
\email{Corresponding author: huiyang.physics@gmail.com}
\affiliation{Department of Physics and Astronomy, Johns Hopkins University, Baltimore, Maryland 21218, USA}
\affiliation{Department of Physics and Astronomy, University of Pittsburgh, Pennsylvania 15213, USA}

\begin{abstract}
We propose a minimal model starting from a parent Chern band with quartic dispersion that can describe the spin-valley polarized electrons in rhombohedral tetralayer graphene. The interplay between repulsive and attractive interactions on top of that parent Chern band is studied. We conduct standard self-consistent mean-field calculations, and find a rich phase diagram that consists of metal, quantum anomalous Hall crystal, chiral topological superconductor, as well as trivial gapped Bose--Einstein condensate. In particular, there exists a topological phase transition from the chiral superconductor to the Bose--Einstein condensate at zero temperature. Motivated by the recent experimental and theoretical studies of composite Fermi liquid in rhombohedral stacked multilayer graphene, we further generalize the physical electron model to its composite fermion counterpart based on a field theory analysis. The chiral superconductor phase of the composite fermion becomes the nonabelian Moore--Read quantum Hall phase. We argue that a chiral (pseudo-)spin liquid phase can emerge in the vicinity of this Moore--Read quantum Hall phase. Our work suggests rhombohedral multilayer graphene as a potential platform for rich correlated topological phases. 
\end{abstract}

\maketitle

\section{Introduction} 

Topological superconductor (TSC) is one of the most attractive concepts in modern condensed matter physics. It can host nonabelian quasiparticles, which are the potential building blocks of quantum computation~\cite{Read2000paired,Schnyder2008Classification, Nayak2008Non,Kitaev2009Periodic,ZhangRMP,Sato2017}. 
In the past decades~\cite{Sato2017,Hsu2017Topological, Le2024Superconducting,Yin2022Topological,Mielke2022Time,Xu2024Observation,Wan2024Unconventional,Ming2023Evidence,Nandkishore2012Chiral,Theuss2024Single,Jiao2020Chiral,Hayes2021Multicomponent,Ran2019Nearly,Huy2007Superconductivity,Aoki2001Coexistence,Stewart1984Possibility,Saxena2000Superconductivity,Ghosh2021Thermodynamics,Maeno1994Superconductivity}, experimental efforts have been made in searching for chiral topological superconductors, in which the time reversal symmetry is broken. Recent experimental and theoretical progress in graphene-based two-dimensional materials provide a new platform for unconventional superconductors and correlated insulators~\cite{Cao2018Unconventional,Yankowitz2019Tuning,Lu2019Superconductors,Han2024signatures,Geier2024Chiral,Cai2023Signatures,Zhou2023fractional,dong2024theory,dong2024stability,dong2024anomalous,soejima2024anomalous,Lu2024Fractional,Han2024signatures,Chou2024intravalley,Kim2024Topological,choi2024electric,sarma2024thermal,Han2023Two,Han2024Quantum,Reddy2024non}. While the moir\'e superlattice often plays an important role in those systems~\cite{Cao2018Unconventional,Yang2019effects,Yankowitz2019Tuning,Lu2019Superconductors,Chen2019signatures,Zhou2021superconductivity,Pantaleon2023superconductivity,Zhang2023enhanced,Holleis2023ising,Li2024tunable,Zhang2024twist,Patterson2024superconductivity,Yang2024diverse,lu2024extended,Adarsh2024Ext,xia2023helical,waters2024topological,Yang2024Exciton,Song2024Phase,Jia2024Moire,Jonah2024Moire,kwan2023moire,yu2024moire,waters2024inter,wang2025fractional}, it has been shown that the rhombohedral stacked multilayer graphene exhibits chiral TSC and gate-tunable Chern bands in the absence of moir\'e pattern, where TSC and anomalous Hall crystal (AHC) are observed at the same electron density \cite{Cai2023Signatures,Lu2024Fractional,Han2024signatures,Chou2024intravalley,Kim2024Topological,Geier2024Chiral}.

The intertwinement of topology and correlations in graphene-based two-dimensional materials can lead to rich physics. Motivated by this, we propose a minimal model starting from a parent Chern band with quartic dispersion found in rhombohedral tetralayer graphene~\cite{Han2024signatures}. We consider the interplay between repulsive and attractive interactions among electrons. As a concrete example, we choose the Coulomb interaction as the repulsive interaction and the electron-phonon coupling as the attractive one, two of which are the most commonly discussed interactions in solids. Note that the formation of the anomalous Hall crystal has been proposed in Ref.~\cite{Dong2025phonon}, breaking the translation symmetry explicitly and providing the possibility for electron-phonon coupling. Our self-consistent mean-field calculations show a rich phase diagram consisted of metal, AHC, chiral TSC, as well as the trivial gapped Bose-Einstein condensate (BEC). In particular, we show there exists a topological phase transition between TSC and BEC phases at zero temperature. Motivated by the recent experimental and theoretical studies of composite Fermi liquid~\cite{Lu2024Fractional,Zhou2023fractional,zhang2024vortex,dong2024anomalous,dong2024stability,dong2024theory,soejima2024anomalous}, we further provide a field theory analysis that extend our investigation for physical electrons moving under the Berry curvature to composite fermions coupled to dynamical gauge fields, and argue that a chiral (pseudo-)spin liquid (CSL) phase~\cite{Kalmeyer1987,Kalmeyer1989,Savary2016quantum,zhou2017quantum,wang2022structure,kitaev2006anyons,Yao2007Exact,Yao2011Fermionic,szasz2020chiral} can emerge in the vicinity of the chiral TSC phase of the composite fermion~\cite{Son:2015}, i.e., the nonabelian Moore--Read quantum Hall phase~\cite{Moore1991MR}. Our work sheds light on investigating various correlated topological phases in graphene-based materials.

\section{The model}

We consider the following minimal model that captures the spin-valley polarized electrons projected to a single parent band,
\beq
H = H_0 + H_{\rm int}.\label{Eq:Hamiltonian}
\eeq
Here, $H_0 = \sum_{\bm k} \epsilon({\bm k}) c^\dagger_{\bm k} c_{\bm k}$ is the kinetic term, where $c_{\bm k}^\dagger$ is the creation operator of an electron with momentum ${\bm k}$ in the parent band. According to the spin-valley polarization, the electron is single-flavored. In rhombohedral tetralayer graphene, the parent band is believed to be a Chern band with quartic dispersion $\epsilon({\bm k}) =  \alpha |{\bm k}|^4$~\cite{Han2024signatures}, and the Brillouin zone (BZ) is chosen to be hexagonal to respect the $C_3$ symmetry. The electron filling is fixed to $\nu=1$. The interaction $H_{\rm int}$ can be divided into the repulsive part and the attractive part. As an example, we consider the repulsive interaction from the electron-electron Coulomb interaction $H_\text{e-e}$ and the attractive interaction from the electron-phonon coupling $H_\text{e-ph}$~\cite{Tan2024parent}, both of which are projected to the parent Chern band,
\begin{subequations}
	\begin{align}
		H_\text{e-e} &= \sum_{{\bm k}_1 {\bm k}_2 {\bm k}_3 {\bm k}_4} \tilde V_{{\bm k}_1 {\bm k}_2 {\bm k}_3 {\bm k}_4} c^\dagger_{{\bm k}_1} c^\dagger_{{\bm k}_2} c_{{\bm k}_3} c_{{\bm k}_4}, \\
		H_\text{e-ph} &= -\sum_{{\bm k}_1 {\bm k}_2 {\bm k}_3 {\bm k}_4} \tilde U_{{\bm k}_1 {\bm k}_2 {\bm k}_3 {\bm k}_4} c^\dagger_{{\bm k}_1} c^\dagger_{{\bm k}_2} c_{{\bm k}_3} c_{{\bm k}_4},  
	\end{align}
\end{subequations}
with $\tilde V_{{\bm k}_1 {\bm k}_2 {\bm k}_3 {\bm k}_4} = v V({{\bm k}_1 - {\bm k}_4}) {\mathcal F}_{{\bm k}_1,{\bm k}_4} {\mathcal F}_{{\bm k}_2,{\bm k}_3} \delta_{{\bm k}_1 + {\bm k}_2 - {\bm k}_3 - {\bm k}_4}$ and $\tilde U_{{\bm k}_1 {\bm k}_2 {\bm k}_3 {\bm k}_4} =u U({{\bm k}_1 - {\bm k}_4}) {\mathcal F}_{{\bm k}_1,{\bm k}_4} {\mathcal F}_{{\bm k}_2, {\bm k}_3}\delta_{{\bm k}_1 + {\bm k}_2 - {\bm k}_3 -{\bm k}_4}$. Here $u,v$ are positive coupling constants that denote the interaction strengths. The electron-electron interaction is given by the gate-screened Coulomb potential $V({\bm q}) = \tanh{d |{\bm q}|} / (d |{\bm q}|)$, while the electron-phonon interaction takes the form of $U({\bm q}) = q_0^2 /(|{\bm q}|^2 + q_0^2)$~\footnote{Note that, this form of electron-phonon interaction asymptotically approaches the delta function as $q_0\rightarrow 0$. This potential is taken ad hoc and not derived from first principle.}. In this work, we fix $d=10.0/|{\bm K}|$ at the order of the screening length~\cite{Zhou2023fractional,Tan2024parent}, where ${\bm K}$ denotes the momentum between the nearest two inequivalent $K$-points of the hexagonal BZ. We then define the energy scale for kinetic term $H_0$ as $\epsilon_K=\alpha |{\bm K}|^4$, and evaluate $(u,v)$ in terms of $\epsilon_K$. The projected form factor ${\mathcal F}$ encodes the quantum geometry of the parent band~\cite{Tan2024parent},
\begin{equation}
    {\mathcal F}_{{\bm k},{\bm k}^\prime} = \exp \bigg{[} -\frac{{\mathcal B}_0}{4}(|{\bm k}^{\prime} - {\bm k}|^2 + 2i {\bm k}^\prime \times {\bm k}) \bigg{]}.
\end{equation}
Here, ${\mathcal B}_0$ is the Berry curvature which can be taken as fixed parameter. Since the quartic parent Chern band can be approximately viewed as the lowest Landau level, in the following we assume a uniform Berry curvature ${\mathcal B}_0 = 2\pi /\Omega_{\rm BZ}$ corresponding to Chern number equal to one for the parent band, where $\Omega_{\rm BZ}$ is the area of the first BZ.

{\it Self-consistent calculation}--- The competition between $H_\text{e-e}$ and $H_\text{e-ph}$ motivates us to investigate the full phase diagram in terms of $u$ and $v$. To see this, we conduct self-consistent mean-field calculations for the model Hamiltonian Eq.~(\ref{Eq:Hamiltonian}). We start from both particle-particle and particle-hole mean-field channels, and seek for self-consistent solutions for different $(u,v)$. For simplicity, we define $S({\bm q}) = v V({\bm q}) - u U({\bm q})$, such that:
\begin{equation}
    H_{\rm int} = \sum_{{\bm k}_1 {\bm k}_2 {\bm k}_3 {\bm k}_4} \tilde S_{{\bm k}_1 {\bm k}_2 {\bm k}_3 {\bm k}_4} c^\dagger_{{\bm k}_1} c^\dagger_{{\bm k}_2} c_{{\bm k}_3} c_{{\bm k}_4}, 
\end{equation}
where $\tilde S_{{\bm k}_1 {\bm k}_2 {\bm k}_3 {\bm k}_4} = S({\bm k}_1 - {\bm k}_4) {\mathcal F}_{{\bm k}_1, {\bm k}_4} {\mathcal F}_{{\bm k}_2, {\bm k}_3} \delta_{{\bm k}_1 + {\bm k}_2 - {\bm k}_3 - {\bm k}_4}$. 

For the particle-hole channel, we decompose the electron momentum as $({\bm k} + {\bm g}_i)$, where ${\bm k}$ is defined in the first BZ, and ${\bm g}_i$ labels the reciprocal lattice vectors under $C_3$ rotation. Under the Hartree--Fock approximation~\cite{Tan2024parent}, the self-consistent mean-field Hamiltonian can be divided into the Hartree term with ${\bm k}_1 = {\bm k}_4$ and ${\bm k}_2 = {\bm k}_3$,
\begin{equation}
    H_{\rm H} = \sum_{\substack{{\bm k}_1 {\bm k}_2 \\ {\bm g}_1 {\bm g}_2 {\bm g}_3 {\bm g}_4}} 2 \tilde S_{{\bm k}_1 {\bm k}_2 {\bm k}_3 {\bm k}_4}^{{\bm g}_1 {\bm g}_2 {\bm g}_3 {\bm g}_4}  P_{{\bm g}_1 {\bm g}_4}({\bm k}_1) c^\dagger_{{\bm k}_2 + {\bm g}_2} c_{{\bm k}_2 + {\bm g}_3} + E_{\rm H},
\end{equation}
and the Fock term with ${\bm k}_1 = {\bm k}_3$ and ${\bm k}_2 = {\bm k}_4$,
\begin{equation}
    H_{\rm F} = -  \sum_{\substack{{\bm k}_1 {\bm k}_2 \\ {\bm g}_1 {\bm g}_2 {\bm g}_3 {\bm g}_4}} 2 \tilde S_{{\bm k}_1 {\bm k}_2 {\bm k}_3 {\bm k}_4}^{{\bm g}_1 {\bm g}_2 {\bm g}_3 {\bm g}_4} P_{{\bm g}_1 {\bm g}_3}({\bm k}_1)c^\dagger_{{\bm k}_2 + {\bm g}_2} c_{{\bm k}_2 + {\bm g}_4} + E_{\rm F}.
\end{equation}
Here, $E_{\rm H}$ ($E_{\rm F}$) is the constant energy shift from Hartree (Fock) contribution, while $ \tilde S_{{\bm k}_1 {\bm k}_2 {\bm k}_3 {\bm k}_4}^{{\bm g}_1 {\bm g}_2 {\bm g}_3 {\bm g}_4} = S({\bm k}_1 + {\bm g}_1 - {\bm k}_4 - {\bm g}_4) {\mathcal F}_{{\bm k}_1 + {\bm g}_1, {\bm k}_2 + {\bm g}_4} {\mathcal F}_{{\bm k}_2 + {\bm g}_2, {\bm k}_1 + {\bm g}_3} \delta_{{\bm g}_1 + {\bm g}_2 - {\bm g}_3 - {\bm g}_4}$.

For the particle-particle channel, we assume that the pairing happens between electrons with opposite momenta. In this way, the BdG Hamiltonian reads,
\begin{equation}\label{eq:SC_mean_field}
    H_{\rm BdG} = \sum_{\bm k} \Psi^\dagger_{\bm k} \begin{pmatrix}
        [\epsilon({\bm k}) - \mu]/2 & \Delta({\bm k}) \\
        \Delta^*({\bm k}) & [-\epsilon({\bm k}) + \mu]/2
    \end{pmatrix} \Psi_{\bm k},
\end{equation}
where $\Psi_{\bm k} = (c_{\bm k}, c^\dagger_{- {\bm k}})^T$ is the Nambu spinor. Here $\mu$ is the chemical potential determined by the self-consistent equation of electron number $N$ at filling $\nu = 1$. When $\mu <0$, the BdG Hamiltonian Eq.~\ref{eq:SC_mean_field} describes a BEC phase. The superconducting gap which respects spatial asymmetry $\Delta(-{\bm k}) = -\Delta({\bm k})$ representing the odd parity self-consistent solution of Eq.~(\ref{eq:SC_mean_field}) for the pairing between single-flavored electrons. We also denote $E({\bm k}) = \sqrt{[\epsilon({\bm k}) - \mu]^2/4 + |\Delta({\bm k})|^2}$ as the dispersion of the Bogoliubov band.


\begin{figure*}[t]
\centering 
\includegraphics[width=2\columnwidth]{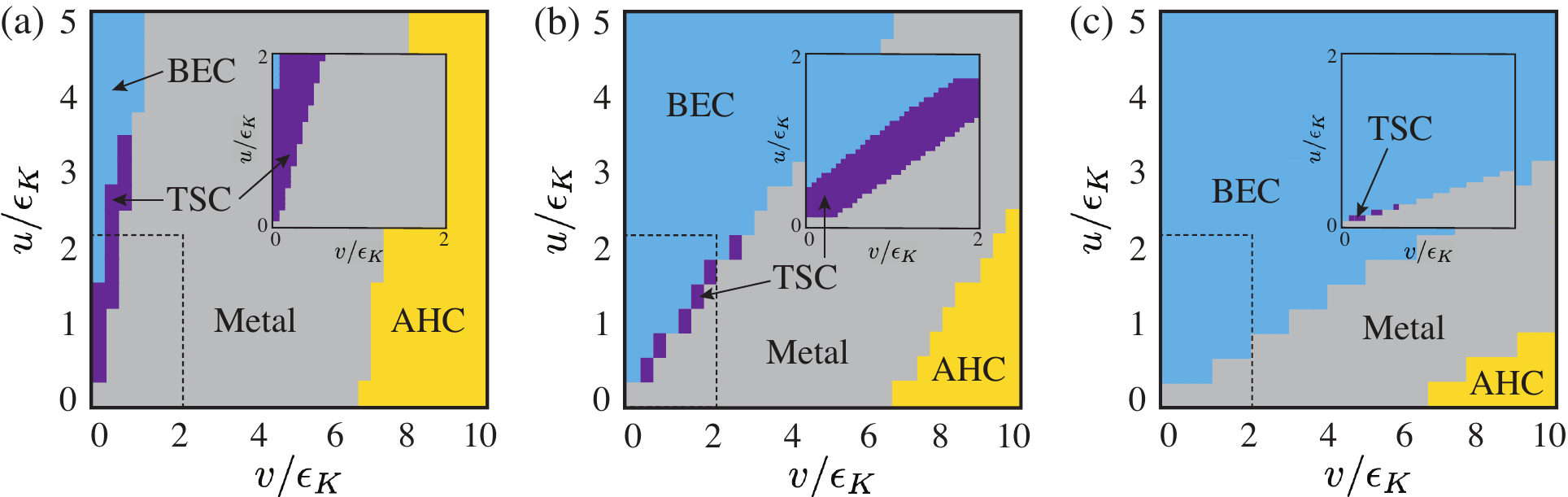}
\caption{Phase diagram of Hamiltonian Eq.~(\ref{Eq:Hamiltonian}) from self-consistent calculations for (a) $q_0=0.05|{\bm K}|$, (b) $q_0=0.3|{\bm K}|$, and (c) $q_0=20|{\bm K}|$. The gray region shows the (semi-)metallic phase. The yellow block stands for the AHC phase. The purple paddles denote the $(p+ip)$-wave TSC, and the blue block represents the trivial gapped BEC phase. The inset zooms in the left bottom corner of the phase diagram with higher numerical resolution.\label{PictureS1}}
\end{figure*}

\section{Phase diagram} 

With above, we conduct self-consistent calculations of Eq.~(\ref{Eq:Hamiltonian}) for the particle-particle and the particle-hole channels numerically and map out the phase diagram Fig.~[\ref{PictureS1}] with $0\le u\le 5\epsilon_K$ and $0\le v\le 10\epsilon_K$. In particular, we vary the parameter $q_0$ representing the range of electron-phonon interaction potential, as Fig.~[\ref{PictureS1}.(a)] $q_0=0.05|{\bm K}|$, Fig.~[\ref{PictureS1}.(b)] $q_0=0.3|{\bm K}|$, and Fig.~[\ref{PictureS1}.(c)] $q_0=20|{\bm K}|$. For $q_0\gg |{\bm K}|$, the interaction becomes nearly uniform, which is close to the BCS limit. On the contrary, when $q_0\ll |{\bm K}|$, the form of interaction asymptotes a delta function, which better describes materials. We note that the phase diagram does not change qualitatively for different $q_0$.

When the values of $u$ and $v$ are approximately comparable, the system favors a (semi-)metallic phase, as shown in the gray region. A strong repulsive Coulomb interaction (i.e., large $v$) can drive the system to an insulating phase, with a finite charge gap open at the first BZ edge, as shown in the yellow region. Within the yellow region, the valence band of $H_{\rm HF}$ has nontrivial Chern number $C=1$ and order parameter $P_{{\bm g}_1{\bm g}_2}({\bm k})$, indicating an AHC phase.

\begin{figure}[t]
\centering 
\includegraphics[width=1\columnwidth]{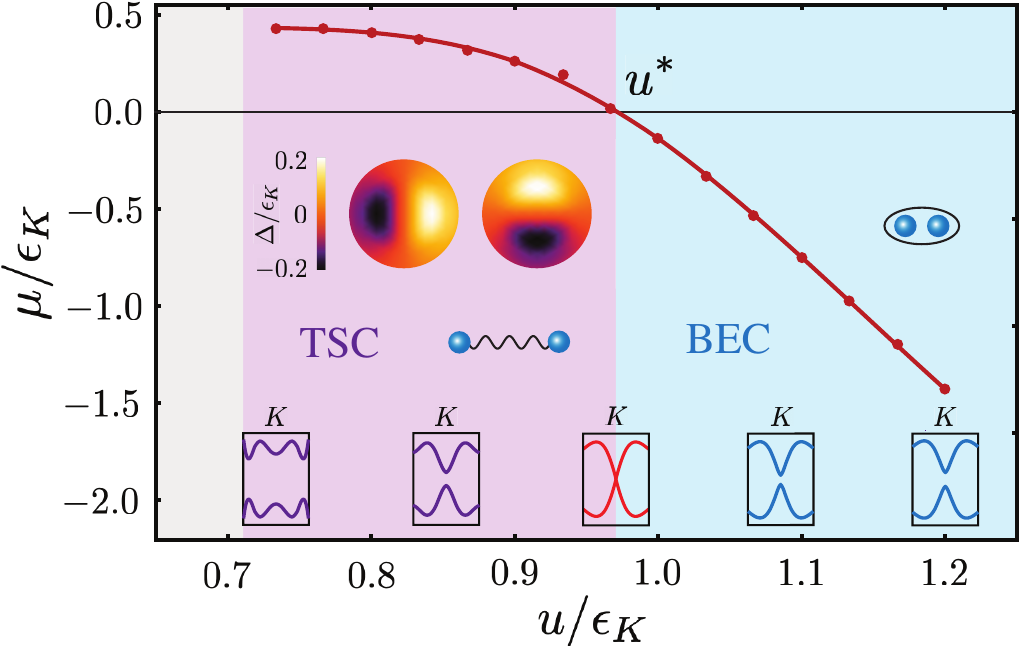}
\caption{\label{Picture4}Transition between the chiral TSC (purple region) and the trivial BEC phase (blue region), with $q_0=0.3|{\bm K}|$, $v = \epsilon_K$ and $0.7\epsilon_K\le u\le 1.2\epsilon_K$. One can see the chemical potential ($\mu$) drop as the increasing of the electron-phonon coupling strength $(u)$. The inset in TSC regime shows superconducting gap distribution with $u= 0.72 \epsilon_K$, $v = \epsilon_K$, where the real and imaginary part of $\Delta({\bm k})$ is shown in the left and right panel, clearly suggesting the feature of a $(p + ip)$-wave superconductor. The bottom insets illustrate the band structure of Bogoliubov quasiparticles. The band gap closes at the critical point $(u^*\approx 0.97 \epsilon_{K},~\mu = 0)$, indicating the topological phase transition from chiral TSC to trivial BEC phase.}
\end{figure}

For the $(u,v)$ in the purple region of Fig.~[\ref{PictureS1}], one can find energetically favored convergent solutions for BdG Hamiltonian Eq.~(\ref{eq:SC_mean_field}). The superconducting gap distribution is shown in the inset of Fig.~[\ref{Picture4}] for $q_0=0.3|{\bm K}|$, whose real (imaginary) part has the profile of $p_x$-wave ($p_y$-wave) pairing, indicating a chiral TSC phase with $(p+ip)$-wave pairing symmetry~\cite{ZhangRMP}\footnote{For the $C_3$ symmetric system, the $p$-wave and the $d$-wave superconductors are related by a time reversal transformation, since the former one has angular momentum $l=+1$ and the latter has $l=+2=-1$ (mod 3).}. The physical origin for the chiral TSC can be understood as follows. For a trivial parent band, due to the Cooper instability, a finite electron-phonon coupling will lead to $s$-wave superconducting phase. For a parent Chern band, the nontrivial Berry curvature will give rise to $p$-wave superconducting phase. This is in accordance with previous work~\cite{Shi2006attractive,Li2018Topological,Simon2022Role}. The vortices in the bulk of this phase can trap nonabelian Majorana zero modes~\cite{ZhangRMP}. The chiral TSC phase exists against variation of $q_0$, though not favored by large $q_0$ as the spatial uniformity of the electron-phonon interaction effectively leads to an enhanced coupling strength of the attractive interaction and hence a trivial gapped BEC phase.

In the blue region of Fig.~[\ref{PictureS1}] we have $\mu <0$ that stands for the BEC phase. The transition between TSC and BEC phases here is rather interesting. The TSC phase has positive chemical potential $\mu$. Deep in the TSC phase, the electrons are dominantly paired close to the Fermi surface, in accordance with the fact that the Bogoliubov band gap is approximately minimized near the Fermi wavevector deep in the TSC phase. This can be checked for the inset of $u = 0.72\epsilon_K$ in Fig.~[\ref{Picture4}] (plotted at $q_0=0.3|{\bm K}|$) showing the corresponding Bogoliubov band dispersion. The Cooper pair is extended in real space. The increasing of electron-phonon interaction $u$ leads to the drop of $\mu$ as well as the decreasing of the Cooper pair size, such that the electrons deep in the Fermi sea start to pair. Different from the case of $s$-wave, for $(p+ip)$-wave there exists a critical $u^*$ where $\mu =0$ and Bogoliubov band gap closes at ${\bm k}=0$. This can be seen that $E({\bm k} = 0) = 0$ when $\mu =0$ and $\Delta({\bm k} = 0) = 0$ for $(p+ip)$-wave. The Chern number for the Bogoliubov band changes from $1$ in TSC phase to $0$ in BEC phase, signaling a topological phase transition from the TSC to the trivial gapped BEC phase. In the BEC phase, the more negative $\mu$ is, the more electrons in the Fermi sea are paired and the size of the Cooper pairs is smaller. The gap of Bogoliubov band is minimal at ${\bm k}=0$, as shown in the insets with $u = 1.08 \epsilon_K$ and $u = 1.2 \epsilon_K$ in Fig.~[\ref{Picture4}]. Details of the topological phase transition from TSC to BEC phases with fixed $v = \epsilon_K$ and varying $u$ are shown in Fig.~[\ref{Picture4}]. We point out that at finite temperature such a phase transition will change to a crossover~\cite{Crepel2023topological}.

\begin{figure}[t]
\centering 
\includegraphics[width=0.95\columnwidth]{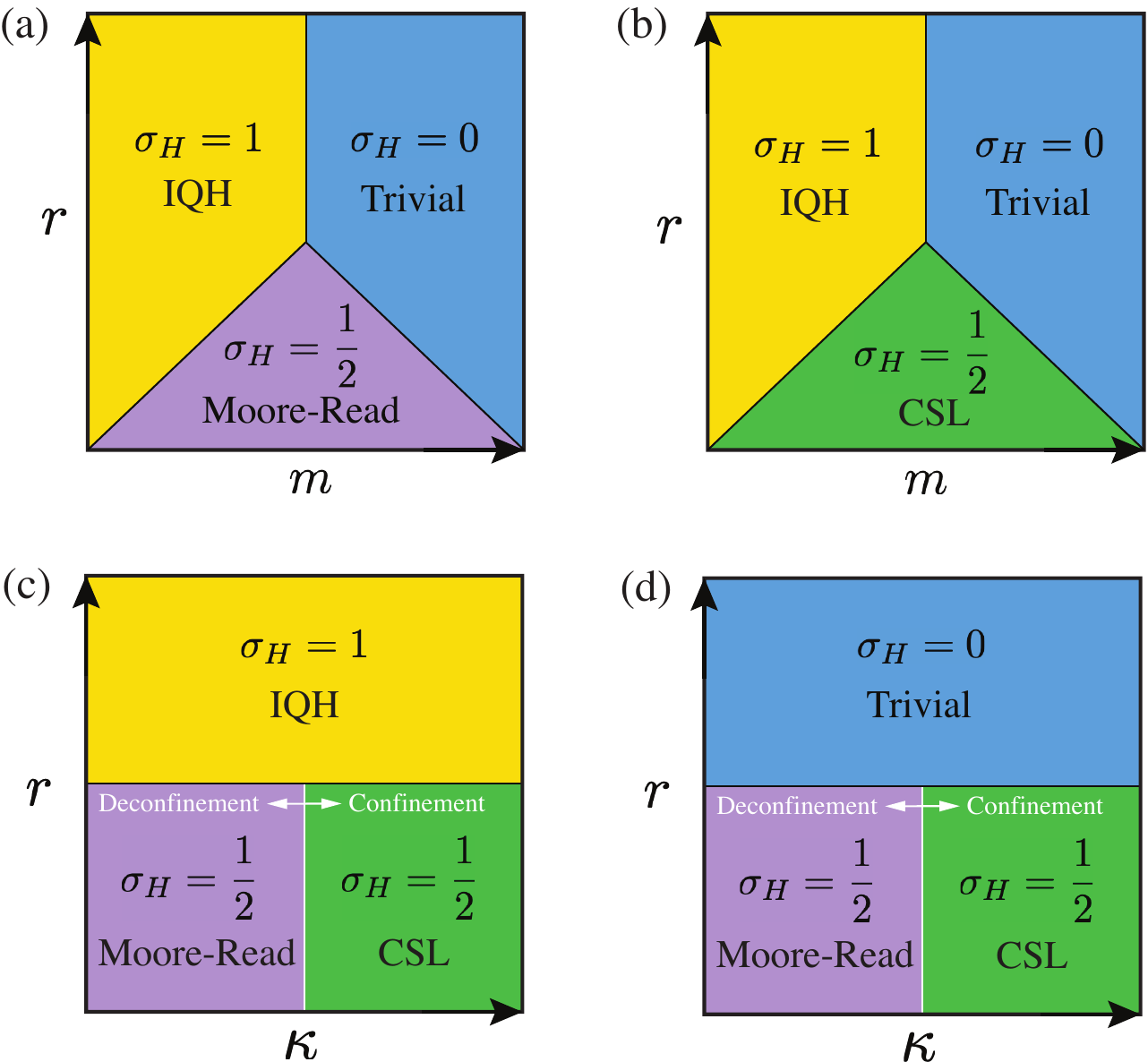}
\caption{\label{Picture1} Schematic phase diagrams for the composite fermion theory Eq.~(\ref{eq:CF}) under fixed $\lambda,t>0$, with (a) small Maxwell coupling $\kappa$, (b) large Maxwell coupling $\kappa$, (c) negative composite fermion mass $m$, (d) positive composite fermion mass $m$.}
\end{figure}

\section{Generalization to composite fermion theory} 

The phase diagram Fig.~[\ref{PictureS1}] provides insights to study the potential nonabelian quantum Hall phase in rhombohedral multilayer graphene, in which the composite Fermi liquid has been studied both experimentally~\cite{Lu2024Fractional} 
and theoretically~\cite{Zhou2023fractional}. It is natural to extend our model Eq.~(\ref{Eq:Hamiltonian}) with physical electrons moving in the background Berry curvature to that of the composite fermion coupled to the dynamical gauge field. The chiral TSC phase for the composite fermion is proposed to be described by the Moore--Read~\cite{Moore1991MR,Read2000paired} quantum Hall state, which exhibits nonabelian topological order. Below we generalize the phase diagram Fig.~[\ref{PictureS1}] into the composite fermion case with the help of field theory. The composite fermions in the presence of pairing can be modeled as~\cite{Son:2015,Seiberg2016,Senthil:2018a}
\beq
\mathcal{L}&=&\bar{\psi}i\gamma^\mu(\partial_\mu -ia_\mu)\psi-\frac{i}{8\pi}a\mathrm{d}a+m\bar{\psi}\psi+\frac{1}{2\kappa}(\mathrm{d}a)^2\nn\\
&&+|D_{2a}\Phi |^2+r |\Phi |^2+t |\Phi |^4-\lambda \left(\Phi^\dag \psi_+\psi_- +h.c.\right)\nn\\
&&-\frac{i}{2\pi}a\mathrm{d}b-\frac{2i}{4\pi}b\mathrm{d}b-\frac{i}{2\pi}b\mathrm{d}A.\label{eq:CF}
\eeq
Here gamma matrices can be chosen as $\gamma^0=\sigma^z$, $\gamma^1=i\sigma^y$, and $\gamma^2=-i\sigma^x$~\cite{Son:2015}. The composite fermion $\psi$ is a two-component Dirac fermion with mass $m$. It is charged under the dynamical compact $U(1)$ gauge field $a$, and $\kappa$ is the Maxwell coupling of $a$. The complex boson field $\Phi$ describes the pairing of the composite fermion, which is consequently charge-2 under the gauge field $a$. Its condensation is controlled by $r$, with the coupling constant of the quartic term $t$ set to be large and positive. The $\lambda$-term ($\lambda>0$) describes $s$-wave pairing for the Dirac composite fermion, which is equivalent to the chiral TSC for the nonrelativistic composite fermion~\cite{Son:2015}. This is due to the intrinsic Berry curvature of the Dirac fermion, with a similar mechanism that Berry curvature can turn the $s$-wave superconductor into a chiral one in the physical fermion case~\cite{Shi2006attractive,Antonenko2024Making}. Dynamical compact $U(1)$ gauge field $b$ is coupled to the background gauge field $A$ through a mutual Chern--Simons (CS) term, whose flux corresponds to the current of the physical electron. The Maxwell coupling of $b$ is neglected since it is more irrelevant than the self CS term.

The phase diagram of Eq.~(\ref{eq:CF}) can be analyzed at the mean-field level. We first consider the nonpairing phases. This corresponds to $r>0$, where $\Phi$ is gapped and $\psi$ can be integrated out. The resulting topological part of the Lagrangian reads
\begin{equation}\label{eq:top}
    \mathcal{L}_\mathrm{QH}=-\Theta(-m)\frac{i}{4\pi}a\mathrm{d}a-\frac{i}{2\pi}a\mathrm{d}b-\frac{2i}{4\pi}b\mathrm{d}b-\frac{i}{2\pi}b\mathrm{d}A.
\end{equation}
The topological data, including ground state degeneracy, Hall conductance, anyon charge and statistics can be extracted from the standard $K$-matrix theory~\cite{Wen1995}. We find that the Hall conductance is $\sigma_H=0$ or $\sigma_H=1$ for $m$ positive or negative, corresponding to a trivial or an integer quantum Hall (IQH) phase, respectively. Meanwhile, since the absolute value of the determinant of the $K$-matrix for Eq.~(\ref{eq:top}) equals to one, neither of the two phases are topological ordered due to the trivial ground state degeneracy. Thus the $\sigma_H=1$ ($\sigma_H = 0$) phase is topological equivalent to the AHC (gapped BEC) phase in the phase diagram Fig.~[\ref{PictureS1}]. Though the gapped BEC phase exhibits superconductivity while in the $\sigma_H=0$ phase the composite fermion does not pair, these two phases are topological equivalent as discussed in the study of proximity effect~\cite{Fu2008,ZhangRMP}.

When $r<0$, $\Phi$ is condensed, driving the system into pairing phase for the composite fermion $\psi$. Correspondingly, the $U(1)$ gauge field $a$ is Higgsed to $\mathbb{Z}_2$. The pairing potential ensures $\psi$ to remain gapped regardless of the sign of $m$, as long as $|m|<|\lambda\left<\Phi\right>|$. Thus, integrating out $\psi$ naively yields the effective Lagrangian
\begin{equation}
    \mathcal{L}_\mathrm{MR}=\frac{1}{2\kappa}(\mathrm{d}a)^2+\frac{2i}{2\pi}a\mathrm{d}\tilde{a}-\frac{i}{2\pi}a\mathrm{d}b-\frac{2i}{4\pi}b\mathrm{d}b-\frac{i}{2\pi}b\mathrm{d}A,\label{eq:MR}
\end{equation}
where $\tilde{a}$ is the Lagrange multiplier introduced such that $a$ is indeed a $\mathbb{Z}_2$ gauge field~\footnote{Integrating out Majorana fermion will generate a gravitational CS term~\cite{Son:2015}, which is not the focus of our study here and omitted in the field theory Eq.~(\ref{eq:MR}). Self CS term of the $\mathbb{Z}_2$ gauge field is not generated since Majorana fermion does not suffer the parity anomaly.}. The Hall conductance in this phase is $\sigma_H=1/2$, and the vortex whose current is coupled to $\tilde{a}$ has charge $Q_\mathrm{v}=1/4$ and self-statistical angle $\theta_\mathrm{v}=\pi/8$. The quasi hole coupled to gauge field $b$ has half charge $Q_\mathrm{h}=1/2$ and semionic self-statistics $\theta_\mathrm{h}=\pi/2$. These topological data exactly match those in the nonabelian Moore--Read quantum Hall state. However, Eq.~(\ref{eq:MR}) only captures the Abelian part of the anyon statistics. To incorporate nonabelian statistics, one needs to include Majorana zero modes arising naturally in the chiral superconductor phase of the composite fermion. One way to include Majorana after integrating out $\psi$ is to substitute the $\tilde{a}\mathrm{d}a$ term by $(\tilde{a}+\eta i\mathrm{d}\eta/4)\mathrm{d}a$~\cite{Hansson:2012}, where $\eta$ describes the Majorana zero mode trapped at the vortex core. This gives rise to the desired nonabelian statistics~\cite{Hansson:2012} in the Moore--Read state. The nonabelian Moore--Read quantum Hall phase is corresponding to $(p+ip)$-wave chiral TSC phase in Fig.~[\ref{PictureS1}]. It is important to note that the Pfaffian and the anti-Pfaffian quantum Hall phases with similar nonabelian statistics can be also achieved by tuning the pairing symmetry of the composite fermion in Eq.~(\ref{eq:CF})~\cite{Son:2015}.

In the composite fermion theory, there is an additional phase without a counterpart in the physical electron model and phase diagram Fig.~[\ref{PictureS1}]. To see this, note that when the composite fermion pairs, the gauge field $a$ is a $\mathbb{Z}_2$ gauge field which can be confined by increasing the Maxwell coupling $\kappa$ through a confinement-deconfinement transition \cite{Fradkin:1979}. When $a$ is confined, the topological part of the theory reads
\beq
\mathcal{L}_\mathrm{CSL}=-\frac{2i}{4\pi}b\mathrm{d}b-\frac{i}{2\pi}b\mathrm{d}A=\frac{1}{2}\frac{i}{4\pi}A\mathrm{d}A,
\eeq
where in passing to the second equality the gapped gauge field $b$ is integrated out, yielding a same Hall conductance $\sigma_H=1/2$ as the Moore--Read phase discussed previously. This phase is a CSL phase with an Abelian semion whose current is coupled to $b$. Therefore, we conjecture the existence of the CSL phase in the vicinity of the Moore--Read quantum Hall phase with Hall conductance $\sigma_H=1/2$ in multilayer graphene systems.

Equipped with above analysis, we can map out the phase diagram of the composite fermion theory Eq.~(\ref{eq:CF}), as shown in Fig.~[\ref{Picture1}]. Fig.~[\ref{Picture1}.(a)] shows when Maxwell coupling $\kappa$ is small, there will be three phases, IQH, Moore--Read, and trivial gapped phases. This is similar to the result in physical electron model shown in Fig.~[\ref{Picture1}.(a)]. When $\kappa$ is strong, the confinement of the ${\mathbb Z}_2$ gauge field will drive the Moore--Read phase into the CSL phase, as shown in Fig.~[\ref{Picture1}.(b)]. The phase boundaries between the pairing (Moore--Read or CSL) and nonpairing (IQH or trivial) phase of composite fermion should be located at $m=\pm\lambda \left<\Phi\right>$ for $r<0$. Figs.~[\ref{Picture1}.(c-d)] are plotted when the composite fermion mass $m <0$ and $m>0$, respectively, which explicitly shows the confinement-deconfinement transition between the CSL and the Moore--Read phases. Their phase boundary can in principle depend on $\kappa$.

\section{Discussion} 

Topological BCS--BEC transitions have recently gained attention in TMD materials~\cite{Crepel2023topological}. In tetralayer graphene, chiral superconductivity has been reported near the finite-temperature BCS--BEC crossover, although the system still predominantly resides on the BCS side~\cite{Han2024signatures}. This is consistent with our zero-temperature results for the BCS--BEC transition. We therefore propose that such a topological BCS--BEC transition could be detected experimentally in the future, for example via ARPES measurements that track the closing and reopening of the spectral gap~\cite{ShenRMP,Zhang2018Observation,Wang2018Evidence}.

Our field theory analysis predicts that the nonabelian Moore--Read quantum Hall phase is neighbored to a chiral pseudospin liquid phase across a confinement-deconfinement transition of $\mathbb{Z}_2$ gauge field. Thus, our results suggest that the chiral (pseudo-)spin liquid state may arise in the vicinity of the Moore--Read state in experiments, and vice versa. The field theory provides a solid phase diagram, closely tied to our Hartree--Fock analysis. 
These shed light on the search for exotic topological phases as nonabelian quantum Hall states and quantum spin liquids in future numerical simulations and experimental realization.


\section{Acknowledgements} We acknowledge Srinivas Raghu, Jing-Yuan Chen, and Zhaoyu Han for discussions on related topics. We also thank Joel E. Moore, Zhengguang Lu, Oskar Vafek, Chunxiao Liu, Zhihuan Dong,  Taige Wang, Yang-Zhi Chou and Jihang Zhu for helpful discussions. Y.-Q. W. is supported by JQI postdoctoral fellowship at the University of Maryland. Z.-Q. G. is funded through the U.S. Department of Energy, Office of Science, Office of Basic Energy Sciences, Materials Sciences and Engineering Division under Contract No. DE-AC02-05-CH11231 (Theory of Materials program KC2301).

\bibliography{chiral}

\noindent

\onecolumngrid


\appendix

\section{Hartree--Fock calculation for insulating phase}

For the insulating phase we decompose the electron momentum as $({\bm k}+{\bm g})$, where ${\bm k}$ is defined in the first BZ, and ${\bm g}$ labels the reciprocal lattice vectors. Under this notation the Hamiltonian reads
\beq
H_0=\sum_{{\bm k} \in {\rm BZ}} \sum_{{\bm g}} \epsilon({\bm k} + {\bm g}) c^\dagger_{{\bm k} + {\bm g}}c_{{\bm k} + {\bm g}},
\eeq
and
\beq
H_{\rm int} = \sum_{{\bm k} \in {\rm BZ}} \sum_{{\bm g}_1,{\bm g}_2, {\bm g}_3, {\bm g}_4}&& S({\bm k}_1 + {\bm g}_1 - {\bm k}_4 - {\bm g}_4) {\mathcal F}_{{\bm k}_1 + {\bm g}_1, {\bm k}_4 + {\bm g}_4}{\mathcal F}_{{\bm k}_2 + {\bm g}_2, {\bm k}_3 + {\bm g}_3} c^\dagger_{{\bm k}_1 + {\bm g}_1} c^\dagger_{{\bm k}_2 + {\bm g}_2} c_{{\bm k}_3 + {\bm g}_3} c_{{\bm k}_4+ {\bm g}_4}\nn\\
&&\times  \delta_{{\bm k}_1 + {\bm g}_1 +  {\bm k}_2 + {\bm g}_2 - {\bm k}_3 - {\bm g}_3 - {\bm k}_4 - {\bm g}_4}.
\eeq
Under the Hartree--Fock approximation, the interaction Hamiltonian can be decomposed into the Hartree term with ${\bm k}_1 = {\bm k}_4$ and ${\bm k}_2 = {\bm k}_3$:
\beq
H_\mathrm{H} =E_\mathrm{H}+ \sum_{{\bm k} \in {\rm BZ}} \sum_{{\bm g}_1,{\bm g}_2,{\bm g}_3,{\bm g}_4}&& 2S({\bm g}_1 - {\bm g}_4) {\mathcal F}_{{\bm k}_1 + {\bm g}_1, {\bm k}_1 + {\bm g}_4} {\mathcal F}_{{\bm k}_2 + {\bm g}_2 , {\bm k}_2 + {\bm g}_3} P_{{\bm g}_1,{\bm g}_4}({\bm k}_1) c^\dagger_{{\bm k}_2 + {\bm g}_2} c_{{\bm k}_2 + {\bm g}_3} \delta_{{\bm g}_1 + {\bm g}_2 - {\bm g}_3 - {\bm g}_4},
\eeq
and the Fock term with ${\bm k}_1 = {\bm k}_3$ and ${\bm k}_2 = {\bm k}_4$:
\beq
H_\mathrm{F} =E_\mathrm{F} -\sum_{{\bm k} \in {\rm BZ}} \sum_{{\bm g}_1,{\bm g}_2,{\bm g}_3,{\bm g}_4}&& 2S({\bm k}_1 + {\bm g}_1 - {\bm k}_2 - {\bm g}_4) {\mathcal F}_{{\bm k}_1 + {\bm g}_1, {\bm k}_2 + {\bm g}_4}{\mathcal F}_{{\bm k}_2 + {\bm g}_2, {\bm k}_1 + {\bm g}_3} P_{{\bm g}_1 {\bm g}_3}({\bm k}_1) c^\dagger_{{\bm k}_2 + {\bm g}_2} c_{{\bm k}_2 + {\bm g}_4}\delta_{{\bm g}_1 + {\bm g}_2 - {\bm g}_3 - {\bm g}_4}.\quad
\eeq
Here 
\beq
P_{{\bm g}_i,{\bm g}_j}({\bm k}) = \left<  c^\dagger_{{\bm k} + {\bm g}_i} c_{{\bm k} + {\bm g}_j}\right>
\eeq
is the mean-field ansatz. The constant energy shift $E_{\mathrm{H/F}}$ reads
\beq
E_\mathrm{H}&=&-\sum_{{\bm k} \in {\rm BZ}} \sum_{{\bm g}_1,{\bm g}_2,{\bm g}_3,{\bm g}_4} S({\bm g}_1 - {\bm g}_4) {\mathcal F}_{{\bm k}_1 + {\bm g}_1, {\bm k}_1 + {\bm g}_4} {\mathcal F}_{{\bm k}_2 + {\bm g}_2 , {\bm k}_2 + {\bm g}_3} P_{{\bm g}_1,{\bm g}_4}({\bm k}_1) P_{{\bm g}_2,{\bm g}_3}({\bm k}_2)\delta_{{\bm g}_1 + {\bm g}_2 - {\bm g}_3 - {\bm g}_4},\\
E_\mathrm{F}&=&\sum_{{\bm k} \in {\rm BZ}} \sum_{{\bm g}_1,{\bm g}_2,{\bm g}_3,{\bm g}_4} S({\bm k}_1 + {\bm g}_1 - {\bm k}_2 - {\bm g}_4) {\mathcal F}_{{\bm k}_1 + {\bm g}_1, {\bm k}_2 + {\bm g}_4}{\mathcal F}_{{\bm k}_2 + {\bm g}_2, {\bm k}_1 + {\bm g}_3} P_{{\bm g}_1 {\bm g}_3}({\bm k}_1)P_{{\bm g}_2 {\bm g}_4}({\bm k}_2)\delta_{{\bm g}_1 + {\bm g}_2 - {\bm g}_3 - {\bm g}_4}.
\eeq
In calculation we choose a total of 7 BZs with the first BZ centered and the other 6 BZs packed around it, and 108 ${\bm k}$ points within each BZ. The electron filling is set to be $\nu =1$, which means the total number of electron $N$ is equal to the number of ${\bm k}$ points in the first BZ, i.e. $N=108$. The ground state energy of the insulating phase under the Hartree--Fock approximation is 
\beq
E_\mathrm{HF}=\sum_{{\bm k}\in \mathrm{BZ}}\bra{{\bm k}}H_0+H_\mathrm{H}+H_\mathrm{F}\ket{{\bm k}}=-E_\mathrm{H}-E_\mathrm{F}+\sum_{{\bm k}\in \mathrm{BZ}}\epsilon({\bm k}).
\eeq

\section{Mean-field calculation for superconducting phase}

For the superconducting phase we assume the pairing happens between electrons with opposite momenta. The mean-field interaction Hamiltonian reads
\beq
H_\mathrm{SC}=\sum_{{\bm k},{\bm k}^\prime}S({\bm k}-{\bm k}^\prime)\mathcal{F}_{{\bm k},{\bm k}^\prime}^2c^\dagger_{{\bm k}}c^\dagger_{-{\bm k}}\left<c_{-{\bm k}^\prime}c_{{\bm k}^\prime}\right>+h.c.
\eeq
Define the superconducting gap
\beq
\Delta({\bm k})=\sum_{{\bm k}^\prime}\frac{1}{2}\left[S({\bm k}-{\bm k}^\prime)\mathcal{F}_{{\bm k},{\bm k}^\prime}^2-S(-{\bm k}-{\bm k}^\prime)\mathcal{F}_{-{\bm k},{\bm k}^\prime}^2\right]\left<c_{-{\bm k}^\prime}c_{{\bm k}^\prime}\right>,
\eeq
which respects spatial asymmetry $\Delta(-{\bm k})=-\Delta({\bm k})$ for the pairing of spinless fermions. Then the BdG Hamiltonian reads
\beq
H_\mathrm{BdG}=\sum_{\bm k}\Psi_{\bm k}^\dagger
\begin{pmatrix}
    [\epsilon({\bm k})-\mu]/2 & \Delta({\bm k})\\
    \Delta^*({\bm k}) & [-\epsilon({\bm k})+\mu]/2
\end{pmatrix}\Psi_{\bm k},
\eeq
where $\Psi_{\bm k}=(c_{\bm k}, c_{-\bm k}^\dagger)^\mathbf{T}$ is the Nambu spinor, and $\mu$ is the chemical potential. The self-consistent equations for the gap $\Delta({\bm k})$ and the chemical potential $\mu$ read
\beq
&&\Delta({\bm k})=-\sum_{{\bm k}^\prime}\frac{\Delta({\bm k}^\prime)}{4E({\bm k}^\prime)}\left[S({\bm k}-{\bm k}^\prime)\mathcal{F}_{{\bm k},{\bm k}^\prime}^2-S({\bm k}+{\bm k}^\prime)\mathcal{F}_{-{\bm k},{\bm k}^\prime}^2\right],\\
&&N=\sum_{\bm k}1-\frac{\epsilon({\bm k})}{E({\bm k})},
\eeq
where $E({\bm k})=\sqrt{[\epsilon({\bm k})-\mu]^2/4+|\Delta({\bm k})|^2}$ is the dispersion of the Bogoliubov band, which depends on $\mu$ implicitly. $N=108$ is the total number of electrons introduced in the previous section. The total energy of the superconducting state is 
\beq
E_\mathrm{SC}=\sum_{\bm k}\frac{\epsilon({\bm k})-\mu}{2}-E({\bm k})+\frac{|\Delta({\bm k})|^2}{2E({\bm k})}.
\eeq
In numerics we check the convergence of particle-particle and particle-hole channels for each pair of $(u,v)$, and if both channels converge the energetically favored one is picked.

\section{$K$-matrix calculation and nonabelian statistics in field theory}
As stated in the main text, in the nonpairing phases of the composite fermion, the topological part of the Lagragian reads
\begin{equation}
    \mathcal{L}_\mathrm{QH}=-\Theta(-m)\frac{i}{4\pi}a\mathrm{d}a-\frac{i}{2\pi}a\mathrm{d}b-\frac{2i}{4\pi}b\mathrm{d}b-\frac{i}{2\pi}b\mathrm{d}A.
\end{equation}
Its corresponding $K$-matrix reads
\beq
K_\mathrm{QH}=\begin{pmatrix}
\Theta(-m) & 1 \\ 1 & 2
\end{pmatrix},
\eeq
which does not host topological order since $|\det K|=1$. The charge vector is $\mathbf{p}=(0,1)^\mathbf{T}$, and hence the Hall conductance is $\sigma_H=\mathbf{p}^\mathbf{T}K_\mathrm{QH}^{-1}\mathbf{p}=-\Theta(-m)$. Thus $\sigma_H=0$ for $m>0$ and $\sigma_H=1$ for $m<0$. A topological phase transition occurs at $m=0$, where the composite fermion is gapless.

In the Moore--Read phase, the effective Lagrangian capturing the Abelian part of the anyon statistics is
\begin{equation}
    \mathcal{L}_\mathrm{MR}=\frac{1}{2\kappa}(\mathrm{d}a)^2+\frac{2i}{2\pi}a\mathrm{d}\tilde{a}-\frac{i}{2\pi}a\mathrm{d}b-\frac{2i}{4\pi}b\mathrm{d}b-\frac{i}{2\pi}b\mathrm{d}A.
\end{equation}
Its corresponding $K$-matrix is
\beq
K_\mathrm{MR}=\begin{pmatrix}
0 & -2 & 1 \\ -2 & 0 & 0 \\ 1 & 0 & 2
\end{pmatrix}
\eeq
with charge vector $\mathbf{q}=(0,0,1)^\mathbf{T}$. The Hall conductance in this phase $\sigma_H=\mathbf{q}^\mathbf{T}K_\mathrm{MR}^{-1}\mathbf{q}=1/2$. The vortex carrying ``magnetic charge" under the $\mathbb{Z}_2$ gauge field $a$ has its current coupled to $\tilde{a}$. Thus, its $l$-vector is $\mathbf{l}_\mathrm{v}=(0,1,0)$. The vortex carries charge $Q_\mathrm{v}=\mathbf{l}_\mathrm{v}^\mathbf{T}K_\mathrm{MR}^{-1}\mathbf{q}=1/4$ and self-statistical angle $\theta_\mathrm{v}=\pi\mathbf{l}_\mathrm{v}^\mathbf{T}K_\mathrm{MR}^{-1}\mathbf{l}_\mathrm{v}=\pi/8$. The quasi hole whose current is coupled to gauge field $b$ has $\mathbf{l}_\mathrm{h}=(0,0,1)$. Therefore it has half charge $Q_\mathrm{h}=\mathbf{l}_\mathrm{h}^\mathbf{T}K_\mathrm{MR}^{-1}\mathbf{q}=1/2$ and semionic self-statistics $\theta_\mathrm{h}=\pi\mathbf{l}_\mathrm{h}^\mathbf{T}K_\mathrm{MR}^{-1}\mathbf{l}_\mathrm{h}=\pi/2$. The incorporation of Majorana zero modes reads
\beq
\mathcal{L}_\mathrm{Majorana}=
\frac{2i}{2\pi}\left(\tilde{a}+\frac{1}{4}\eta i\mathrm{d}\eta\right)\mathrm{d}a-i\tilde{a}_\mu j^\mathrm{v}_\mu-ib_\mu j^\mathrm{h}_\mu-\frac{i}{2\pi}a\mathrm{d}b-\frac{2i}{4\pi}b\mathrm{d}b-\frac{i}{2\pi}b\mathrm{d}A,\label{eq:Maj}
\eeq
where $j^\mathrm{v}_\mu$ and $j^\mathrm{h}_\mu$ are the currents of the vortices and the quasi holes, respectively, while the Majorana zero modes are described by $\eta$. Equations of motion for $\tilde{a}$ and $b$ give rise to
\beq
j^\mathrm{v}_\mu=\frac{1}{\pi}\epsilon_{\mu\nu\lambda}\partial_\nu a_\lambda,\qquad j^\mathrm{h}_\mu=-\frac{1}{2\pi}\epsilon_{\mu\nu\lambda}\partial_\nu(2b_\lambda+a_\lambda+A_\lambda).
\eeq
Accordingly, the Majorana theory Eq.~(\ref{eq:Maj}) can be formally written as
\beq
\mathcal{L}_\mathrm{Majorana}=\frac{1}{4}j^\mathrm{v}_\mu\eta i\partial_\mu\eta+\frac{\pi}{8}j^\mathrm{v}\frac{i}{\partial}j^\mathrm{v}+\frac{1}{4}iA_\mu j^\mathrm{v}_\mu+\frac{\pi}{2}j^\mathrm{v}\frac{i}{\partial}j^\mathrm{h}+\frac{\pi}{2}j^\mathrm{h}\frac{i}{\partial}j^\mathrm{h}+\frac{1}{2}iA_\mu j^\mathrm{h}_\mu,
\eeq
after integrating out all the dynamical gauge fields. Here $1/\partial$ is a shorthand of the current-current interaction mediated by gauge fields. The first term describing vortices trapping Majorana zero modes gives rise to the desired nonabelian statistics. The charge and the Abelian self-statistical angle of vortices, $Q_\mathrm{v}=1/4$ and $\theta_\mathrm{v}=\pi/8$ respectively, can also be manifestly read off, and similarly for the quasi holes.

\end{document}